# Coupling between magnon and ligand-field excitations in magnetoelectric $Tb_3Fe_5O_{12}$ garnet

T. D. Kang, E. Standard, K. H. Ahn, and A. A. Sirenko[a]

*Department of Physics, New Jersey Institute of Technology, Newark, New Jersey 07102, USA*

G. L. Carr

*National Synchrotron Light Source, Brookhaven National Laboratory, Upton, New York 11973, USA*

S. Park, Y. J. Choi, M. Ramazanoglu, V. Kiryukhin, and S-W. Cheong

*Rutgers Center for Emergent Materials and Department of Physics and Astronomy, Rutgers University, Piscataway, New Jersey 08854, USA*

[a] sirenko@njit.edu

ABSTRACT

The spectra of far-infrared transmission in $Tb_3Fe_5O_{12}$ magnetoelectric single crystals have been studied in the range between 15 and 100 $cm^{-1}$, in magnetic fields up to 10 T, and for temperatures between 5 and 150 K. We attribute some of the observed infrared-active excitations to electric-dipole transitions between ligand-field split states of $Tb^{3+}$ ions. Anticrossing between the magnetic exchange excitation and the ligand-field transition occurs at the temperature between 60 and 80 K. The corresponding coupling energy for this interaction is 6 $cm^{-1}$. Temperature-induced softening of the hybrid IR excitation correlates with the increase of the static dielectric constant. We discuss the possibility for hybrid excitations of magnons and ligand-field states and their possible connection to the magnetoelectric effect in $Tb_3Fe_5O_{12}$.

75.80.+q, 75.30.Ds, 75.47.Lx

## I. INTRODUCTION

Rare earth iron garnets ($RE$-IG), $RE_3Fe_5O_{12}$, have attracted a lot of attention in the past due to huge magnetostriction, which is directly related to the anisotropy of the crystal field of the $RE^{3+}$ ions,[1-3] where $RE$=Tb, Dy, Ho, Er, *etc*. Recently, it was found that Tb-IG exhibits magnetoelectric and magnetodielectric properties in surprisingly low magnetic fields.[4] Compared to traditional multiferroics, such as $RE$MnO$_3$ and $RE$Mn$_2$O$_5$, where dielectric constant changes in external magnetic fields of a few Tesla,[5,6] Tb-IG shows significant changes in the field of less than 0.2 T. This discovery opens an opportunity for practical applications of $Tb_3Fe_5O_{12}$, especially in the form of thin films that could combine magnetostriction and magnetodielectric properties. However, a number of important questions about the mechanisms that drive magnetodielectric effects at the microscopic level remain unanswered. What is special about Tb ions and why has the same magnetodielectric effect not been observed so far in other similar $RE$-IG compounds with, *e.g.*, $RE$=Dy, Ho, Er? Interaction between the spin, lattice, and electronic structure excitations may provide a clue to understanding the unique properties of $Tb_3Fe_5O_{12}$ in particular and multiferroic effects in general.

Recently, the importance of the low-frequency infrared-active excitations in relationship with the intriguing multiferroic effects has been understood. An electric-dipole excitation called "electromagnon" has been discovered in a number of multiferroics such as $RE$MnO$_3$ and TbMn$_2$O$_5$.[7,8] However, the theoretical picture for electromagnons in multiferroics is still under development.[9,10] For example, the Dzialoshinski-Moriya (DM) interaction between non-collinear spins that allows coupling between magnons and electric field cannot always explain experimental data for polarization selection rules of electromagnon absorption and it is still not clear how two excitations (magnons and optical phonons) interact while being separated by hundred(s) of cm$^{-1}$ on the energy scale. For $RE$-based multiferroics $RE$Mn$_2$O$_5$ we recently proposed an alternative explanation.[11] It is based on the contribution of the forced electric-dipole transitions between the 4$f$ energy levels of the $RE^{3+}$ ions to the static values of the dielectric constant. The exchange interaction between $RE^{3+}$ and Mn magnetic moments results in hybrid excitations: ligand field + magnons. Such excitations were experimentally observed in HoMn$_2$O$_5$. In this scenario, a spontaneous polarization that appears at the temperatures below the ferroelectric phase transition removes $RE^{3+}$ from the center of inversion thus enabling the otherwise forbidden electric-dipole optical transitions between the ligand field states of $RE^{3+}$. This scenario will be verified below for another $RE$-based magnetoelectric system: Tb$_3$Fe$_5$O$_{12}$ garnet.

In this paper, we present a systematic study of the far-IR transmission spectra in Tb$_3$Fe$_5$O$_{12}$ in the vicinity of the critical magnetic fields and the transition temperatures. We will argue that the tuning of the dielectric constant by temperature and magnetic field can be connected to the appearance of the hybrid modes: {magnetic + ligand field electronic excitations of Tb$^{3+}$}.

## II. MATERIAL PROPOERTIES of Tb-IG

The magnetic and crystal structure of $RE$-IG is described in Refs. [[12,13]]. Tb$_3$Fe$_5$O$_{12}$ crystals form a cubic structure with a space group $Ia3d$($O_h^{10}$). Tb$^{3+}$ ions with the ground state $^7F_6$ are in the 24$d$ dodecahedral sites with the local orthorhombic symmetry 222($D_2$). There are several non-equivalent Tb ions in each unit cell with the same surrounding field, but the axes are inclined to each other. This has the overall effect of producing an average cubic symmetry. $Fe^{3+}$ ions occupy two sites: $16a$ octahedral sites with the $\overline{3}$ ($C_{3i}$) symmetry and $24c$ tetrahedral sites with the $\overline{4}$ ($S_4$) symmetry. Below the transition temperature of $T_N$ =550 K, the iron spins are ordered in a ferrimagnetic structure with spins aligned in the [1 1 1] direction. Among six possible exchange interactions between spins in three different magnetic subsystems, only two dominate.[12] The main magnetic superexchange interaction is between Fe in two different sites: spins of Fe in the tetrahedral site are antiparallel to those of the octahedral site. Another important interaction is between Tb and Fe in the tetrahedral site resulting in the Tb spins to be antiparallel to Fe moments in the tetrahedral sites, and, hence, antiparallel to the net magnetic moment of Fe. Below 150 K, a rhombohedral distortion of the cubic cell causes the canting of Tb spins, which is usually described as a "double umbrella structure".[12,14] The symmetry of $Tb^{3+}$ is lowered from 222($D_2$) tetragonal to 2($C_2$) monoclinic. Note that $Tb^{3+}$ is not at the center of inversion that is important for the future discussion of the selection rules for the crystal field transitions.

## III. SAMPLES and MEASUREMENTS

The high-temperature flux growth technique was utilized to produce bulk crystals of $Tb_3Fe_5O_{12}$ (see Ref. [4] for details). Samples with a typical cross section area of about 4×4 $mm^2$ and different thickness in the range between 0.3 and 0.6 mm were used for the light transmission experiments. The incident light beam was perpendicular to the [1 1 1] or [1 0 0] crystallographic planes. The opposite sides of the samples were polished and wedged with a 3° offset in order to suppress interference fringes. Transmission intensity was measured at the National Synchrotron Light Source, Brookhaven National Laboratory, at the U4IR and U12IR beamlines equipped with an Oxford optical cryostat with magnetic field of up to 10 T, Bruker IR spectrometer, and a LHe-pumped (~1.4 K) bolometer. The spectral resolution of 0.3 cm$^{-1}$ was chosen to be significantly smaller than the typical width of the absorption lines of about 2 cm$^{-1}$. Polarization of the transmitted light was scrambled by a light cone and was not analyzed in our experiments. For each sample the raw data of transmitted intensity were normalized to transmission through an empty aperture with the size equal to that of the sample.

## IV. EXPERIMENT at ZERO MAGNETIC FIELD

Figure 1 shows normalized transmission spectra of $Tb_3Fe_5O_{12}$ measured at $T = 5$, 17, and 80 K in a zero external magnetic field. The low-temperature absorption lines appear at 69, 73, and 81 cm$^{-1}$. An additional line at 47 cm$^{-1}$ demonstrated a significant increase of the oscillator strength with the temperature becoming more pronounced for $T>12$ K. A strong decrease of the transmitted intensity above 90 cm$^{-1}$ is due to the absorption by the optical phonons.[15] The transmission intensity map [see Fig. 2(a)] was measured with the temperature increments of 2 K. The frequencies of the three absorption lines at 47, 73, and 81 cm$^{-1}$ show practically no temperature dependence [see Fig. 2(b)]. In contrast, the forth line demonstrates a significant softening from 69 to 30 cm$^{-1}$ in the temperature range between 5 and 140 K.

The assignment process for the observed IR excitations to electric-dipole or magnetic-dipole IR transitions is not straightforward in transmission experiments, especially without a proper light polarization analysis. Among more reliable optical techniques could be variable-incidence-angle polarized reflectivity and, of course, full-Muller matrix ellipsometry.[16] Note, however, that these techniques are not routinely available for the far-IR spectral range yet. Here we will use several indirect evidences, such as temperature and magnetic field dependencies as well as a comparison with the earlier far-IR studies of other *RE*-garnet compounds.[17] For example, strong temperature dependence for the frequency of the line, which appears at 69 cm$^{-1}$ (*T*=5K), is typical for magnetic excitations.[17,18] In contrast, weak temperature dependence is expected for the frequency of the single ion and crystal-field-type excitations, especially far from the phase transitions. The ligand field spectrum should change significantly if one *RE* ion is substituted by another one in *RE*-IG. Indeed, our preliminary IR transmission measurements for a similar compound, $Dy_3Fe_5O_{12}$, show a completely different set of absorption lines at 13, 22, 28, 43, 59, and 75 cm$^{-1}$. Thus, we can attribute the lines at 47 and 73 to ligand-field (LF) transitions of $Tb^{3+}$ ions in Tb-IG. The energies are determined by the combination of the crystalline electric field and the exchange field produced by iron, both nearly temperature-independent since the Fe sublattice remains completely ordered at low temperatures. Detailed information on the crystal field and exchange interactions in *RE*-IG can be found in Ref. [[19]]. The line at 81 cm$^{-1}$ was observed previously in a number of *RE*-IG and had been attributed to the lowest-frequency optical phonon.[17] This interpretation is supported by the typical red shift $\sim \left( m_{RE} \right)^{-\frac{1}{2}}$ of the line frequency for *RE*-IG compounds for different mass $m_{RE}$ of the *RE* ions: from 83 for Sm down to 79 cm$^{-1}$ for Er. Note, however, that this line in Tb-IG is significantly weaker than other IR optical phonons with the frequencies higher than 100 cm$^{-1}$. We did not observe any peaks around 81 cm$^{-1}$ in the Raman spectra of *RE*-IG, so this line cannot be attributed to the Raman-active optical phonon. Thus its interpretation as an IR-active optical phonon should hold.

In addition to phonons and crystal-field excitations, far-IR spectra of ferrimagnetic materials can exhibit magnetic excitations related to the spins of iron and *RE* ions, such as magnons. An acoustic ferrimagnetic mode that corresponds to the strongest superexchange Fe-Fe interaction falls in a very low frequency range. It has been observed in a non-rare-earth yttrium iron garnet in the magnetic field of 0.32 T at ~0.3 cm[-1],[20] which is well below the frequency range for our experiments. The Fe-Tb ferrimagnetic interaction reveals itself in the measured far-IR spectral range. A simplified theory for a *RE*-IG with a collinear spin arrangement was developed in Ref. [[2]]. Although this approach is not directly applicable to the case of Tb-IG with the non-collinear spins, it is helpful for understanding of the major trends in the mode behavior with temperature and magnetic field. If one considers only the interaction between the *RE* and the combined Fe subsystems, then two optical magnetic modes should appear. One is the Kaplan-Kittel (KK) mode $\Omega_M$,[21] which corresponds to the exchange between two magnetic subsystems. Another one $\Omega_{LF}$ corresponds to precession of the *RE* moments in the effective field imposed by the iron magnetization. The zone-center energies of these modes are

$$\Omega_M(T) = \lambda_{ex}\mu_B\left[g_{Fe}M_{Tb}(T) - g_{Tb}M_{Fe}\right]$$
$$\Omega_{LF} = \lambda_{ex}\mu_B g_{Tb}M_{Fe} \tag{1}$$

where $\mu_B$ is the Bohr magneton ($\mu_B \approx 0.4669$ cm[-1]/T), $\lambda_{ex}$ is the exchange constant, $g_{Fe,Tb}$ are the corresponding g-factors, $M_{Tb}$ is the Tb sublattice magnetization, and $M_{Fe}$ is the combined Fe magnetization. At the zone center ($k \approx 0$) the frequency of the $\Omega_{LF}$ mode corresponds to the single ion precession in the magnetic field imposed by Fe on Tb. For an ion with integer $J$, like Tb, the corresponding energy should be equal to the LF splitting of the doubly-degenerated ground state of the free $RE$ ion. Thus, the line at 47 cm$^{-1}$ [Fig. 2(a,b)] can be attributed to $\Omega_{LF}$ in Eq. (1). For $T$<150 K, Tb magnetization $M_{Tb}(T)$ has a strong temperature dependence.[13] In contrast, the iron subsystem magnetization $M_{Fe}$ is almost constant in this temperature range. So, one can expect that $\Omega_M(T)$ will reflect the change of $M_{Tb}(T)$. Note, that the balance between $g_{Fe}M_{Tb}(T)$ and $g_{Tb}M_{Fe}$ determines the temperature trend in $\Omega_M(T)$. If $g_{Fe}M_{RE}(T) < g_{RE}M_{Fe}$, as in Yb-IG, the KK mode frequency increases with $T$ approaching $\Omega_{LF}$ when $RE$ magnetization disappears.[17] In Tb-IG, $g_{Fe}M_{Tb}(T) > g_{Tb}M_{Fe}$ and, hence, the KK mode frequency should decrease with temperature disappearing at the compensation point. This expectation for $\Omega_M(T)$ is supported by the trend in the experimental data shown in Fig. 2(a,b) for the soft mode between 69 and 30 cm$^{-1}$. The KK modes have been studied experimentally in a number of $RE$-IG compounds $RE$=Sm, Ho, Er, Yt (see Ref. [[17]]), but no reports have been done for $RE$=Tb and Dy. Among other factors, the oscillator strength for the KK mode depends on the difference $\left[ g_{Tb} - g_{Fe} \right]^2$ between $g$-factors for Tb and Fe.[2,17] This rule has been confirmed for another compound, Gd-IG, where no KK magnon was detected between 10 and 100 cm$^{-1}$ due to a close proximity of $g$-factors for Gd and Fe: $g_{Gd} \approx g_{Fe} = 2$. The same argument explains why no KK mode was ever detected for the antiferromagnetic interaction between Fe moments in the tetrahedral and octahedral sites in iron garnets using the far-IR experiments.

At low temperature the ratio of $\Omega_M / \Omega_{LF} = 69/47 \approx 1.5$, which allows us to estimate the $g$-factor value of Tb using Eq. (1): $g_{Tb} = g_{Fe} M_{Tb} / [M_{Fe}(1 + \Omega_M / \Omega_{LF})]$. The experimental values for Tb and the total Fe sublattice magnetizations $M_{Tb} \approx 45\mu_B$ / mole and $M_{Fe} \approx 5\mu_B$ / mole,[1,13] and $g_{Fe} = 2$ result in the effective value of $g_{Tb} \approx 7.2$, which is significantly larger than the free-ion value for Tb: $g_0 = 1.5$. Note that such a difference is common in $RE$-IG, where the $J$-mixing and re-population of the crystal-field states at low temperature modify the effective $g$-factor, making it strongly anisotropic. For example, the experimental values for $g_{Dy}$ ($g_x = 11.07$, $g_y = 1.07$, and $g_y = 7.85$) are significantly larger than the corresponding free-ion value of $g_0 = 4/3$ for Dy in Dy-IG.[1]

Unfortunately, the intensities of the magnetic and LF lines decrease dramatically for temperatures above 140 K, where only the optical phonon at 81 cm⁻¹ is visible. This decrease is probably related to the temperature-induced population of the electronic states at 47 and 73 cm⁻¹. Another reason may be related to the disappearance of the rhombohedral distortions for $T > 150$ K that restores the local symmetry of $RE$ ion thus changing the selection rules for the optical transitions between the LF electronic states of Tb³⁺.

An accurate theoretical model is required for description of the magnetic mode spectrum. It should take into account (i) the crystal-field splitting of Tb and its symmetry affected by the rhombohedral distortion of the cubic cell, (ii) variation of the anisotropic $g_{Tb}$ due to the temperature-induced population of the crystal-field levels, and (iii) the anisotropic Tb-Fe exchange for all magnetic sub-systems: two different Tb that correspond to the double-umbrella structure, and two Fe in the tetrahedral and octahedral sites. In this case one could expect an additional splitting of the exchange mode that corresponds to Tb subsystems with the corresponding magnetic moments of $8.18\mu_B$ and $8.9\mu_B$.[1] In our experiment such splitting was not clearly detected. In the following qualitative interpretation of the experimental results, we can continue to relate the soft mode (69 cm$^{-1}$ at low temperature) to the average KK-type exchange excitation, (or a zone-center magnon), and the hard mode (47 cm$^{-1}$) as related to the LF excitation.

In the temperature range around 60 K, an anticrossing occurs between two lower-frequency lines indicating a significant hybridization between the exchange magnon and the LF excitation of Tb$^{3+}$ ions. The results of the fit for the temperature dependence of these two hybrid excitations ($\Omega_{LF-M}^{(1)}$ and $\Omega_{LF-M}^{(2)}$) is shown in Fig. 2(b). The dashed curves are solutions of the LF-M exchange Hamiltonian, which can be written as follows

$$\hat{H} = \begin{bmatrix} \Omega_M & \Delta_{LF-M} \\ \Delta_{LF-M} & \Omega_{LF} \end{bmatrix} \tag{2}$$

in the {|LF>, |M>} basis. Here we assume that $\Omega_{LF}$ and the coupling constant $\Delta_{LF-M}$ are temperature independent, but $\Omega_M$ depends on temperatures. The energy eigenvalues of this Hamiltonian are

$$\Omega_{LF-M}^{(1,2)} = \frac{\Omega_{LF} + \Omega_M}{2} \pm \sqrt{\left(\frac{\Omega_{LF} - \Omega_M}{2}\right)^2 + \Delta_{LF-M}^2} \tag{3}$$

and the corresponding eigenstates are $\left|\Omega_{LF-M}^{(2)}\right\rangle = \begin{pmatrix} \cos\theta \\ \sin\theta \end{pmatrix}$ and $\left|\Omega_{LF-M}^{(1)}\right\rangle = \begin{pmatrix} -\sin\theta \\ \cos\theta \end{pmatrix}$,

where $\theta = \tan^{-1}\left(\sqrt{r^2+1} - r\right)$ and $r = (\Omega_{LF} - \Omega_M)/(2\Delta_{LF-M})$. The best fit for the coupling constant value $\Delta_{LF-M}$ is 6 cm$^{-1}$, which is about 10% of the average energy for the hybrid excitations. With the $\Omega_{LF} - \Omega_M$ separation of approximately $-20$ cm$^{-1}$ and $20$ cm$^{-1}$ at $T$=10 K and 120 K, $\theta$ varies between about 75 and 15 degrees. By calculating $\cos^2(\theta)$ and $\sin^2(\theta)$ versus $T$, as shown in Fig. 2(c), we find that near 10 K the $\left| \Omega_{LF-M}^{(2)} \right\rangle$ state is made of about 93% magnon state and 7 % LF state and vice versa for the $\left| \Omega_{LF-M}^{(1)} \right\rangle$ state. This composition is reversed near 120 K. To understand the nature of the LF-M exchange interaction $\Delta_{LF\text{-}M}$, special theoretical studies would be required. This interaction definitely occurs on the *RE* site. On one hand, the magnetic exchange between Fe and Tb sublattices [Eq.(1)] depends on the ground state of Tb ions, which, on the other hand, is affected by the exchange field produced by the Fe ions influencing the LF energies. We suggest that the specific combination of the KK and LF frequencies, which is determined by the strong Tb magnetization and its *g* factor value, makes Tb-IG unique in the line of other *RE*-IGs.

For a simple antiferromagnetic system with a cubic symmetry and with collinear spins, both the LF and M modes are pure magnetic dipoles. The corresponding optical transitions should be circularly polarized and their spectral weight should contribute to the static value of $\mu(0,T)$ only. However, the strongly anisotropic Fe-Tb exchange, rhombohedral distortion of the lattice, and a possible DM-type of interaction between non-collinear spins of Tb and Fe, can result in a "forced" electric-dipole activity for the coupled LF and M excitations in Tb-IG. Since Tb is not at the center of inversion in Tb-IG, the electric-dipole oscillator strength of the hybrid mode can originate from the "forced" electric-dipole-active LF excitation.[22] In this case, the LF-M modes can have both, the magnetic-dipole and electric-dipole activity. Without a proper analysis of the IR mode polarization, it is hard to decouple the total contribution of the hybrid modes with $\Omega_{LF-M}^{(1,2)}$ energies and $S^{(1,2)}$ oscillator strengths between the changes for the static values of the dielectric $\Delta\varepsilon(0,T)$ and magnetic $\Delta\mu(0,T)$ constants:

$$\Delta\varepsilon(0,T) + \Delta\mu(0,T) = \left(\frac{S^{(1)}}{\Omega_{LF-M}^{(1)}(T)}\right)^2 + \left(\frac{S^{(2)}}{\Omega_{LF-M}^{(2)}(T)}\right)^2.$$

(4)

Figure 3(a,b) shows experimental results for the right hand part of Eq.(4). The experimental data for the combined weight of the hybrid modes were derived from the transmission spectra by using a fit to a multiple oscillator model for magnetic and dielectric functions. Strong variation of the oscillator strength for the line at 47 cm$^{-1}$ in the temperature range between 5 and 30 K [see Fig. 3(a)] is most likely due to the interaction between LF and acoustic excitations. The combined oscillator weight [Fig. 3(b)] is close to the earlier measurement of the static dielectric constant [Fig. 3(c)] from Ref. [4] and the increase with temperature is reproduced in both dependencies. Note that $\varepsilon(0,T)$ in Fig. 3(c) was measured for the direction of electric field [1 -1 0] perpendicular to the light propagation direction in our experiments, which is exactly as it should be for a proper comparison between the static and optical measurements. From the numerical comparison between two curves: $\Delta\varepsilon(0,T) + \Delta\mu(0,T)$ in Fig. 3(b) and $\varepsilon(0,T)$ in Fig. 3(c) one can estimate that the hybrid mode is about (60±30)% electric-dipole active. The error bar takes into account the uncertainty of polarization and normalization errors for transmission measurements. Nevertheless, this comparison allows us to confirm the existence of the ligand-field magnon excitations with a significant electric-dipole activity in Tb-IG. Note that such hybrid LF-M excitation is different from conventional "electromagnons" in multiferroics that are usually attributed in literature to the interaction between magnons and optical phonons.[8] As it will be shown in the next Section, we did not observe any interaction between the hybrid modes and the lowest frequency optical phonon.

## V. EXPERIMENT in MAGNETIC FIELD

A strong magnetic field $H$ applied along [1 1 1] direction (Faraday configuration) causes a linear increase of two LF energies with the slopes of $1\mu_B$ and $7\mu_B$ for transitions at 47 and 73 cm$^{-1}$, respectively [see Fig. 4(a,b)]. Note that the slope of the magnetic field dependence for the LF line at 73 cm$^{-1}$ corresponds to $g = 7$, the same value was estimated for Tb in the previous Section IV [Fig. 4(b)]. As expected, the lowest-frequency optical phonon at 81 cm$^{-1}$ does not change in magnetic field. Even more, the resonance between the phonon and the LF line at ~3 T has no pronounced anticrossing, which indicates their weak interaction. In strong magnetic fields there is a significant softening of the hybrid mode $\Omega_{LF-M}^{(2)}$ with the slope of about $-1.5\mu_B$.

The most interesting part of the magnetic field dependence is in the range for $H <$ 0.5 T [see Fig. 5(a,b)], where the magnetodielectric effect has been previously reported.[4] For both directions of the applied magnetic field, along [1 1 1] and [1 0 0], we observed a non-linear behavior for $\Omega_{LF-M}^{(2)}(H)$. A close zoom on this dependence for $H \parallel$ [1 0 0] is shown in Fig 5(b). Note that the high-field slope for $\Omega_{LF-M}^{(2)}(H)$ increases for this orientation up to $9\mu_B$ compared to that for $H \parallel$ [1 1 1] ($7\mu_B$). Figure 5(b) illustrates that another coupling between the magnon at 69 cm$^{-1}$ and another LF excitation at 73 cm$^{-1}$ (marked with symbol *) occurs at the weak magnetic fields. Two coupled excitations are separated by ~3.5 cm$^{-1}$ in the field range between 0 and 0.5 T. In stronger fields, the Zeeman effect becomes dominant and the LF transition shifts up, while the magnon slowly decreases its energy. Solutions of the following Hamiltonian

$$\widehat{H} = \begin{bmatrix} \Omega_{LF}^* & \Delta_{LF-M}^* & 0 \\ \Delta_{LF-M}^* & \Omega_M & \Delta_{LF-M} \\ 0 & \Delta_{LF-M} & \Omega_{LF} \end{bmatrix} \tag{5}$$

are shown in Fig. 5(b) with blue curves. The zero-field coupling constant $\Delta_{LF-M}^*$ between the higher energy LF and the magnon excitations is estimated to be 2 cm$^{-1}$. The high-field solutions for uncoupled excitations are shown in Fig. 5(b) with dashed lines. The possible hysteresis effect was studied for the ramps of magnetic field up to 10 T and back down to zero. Only minute difference in the oscillator strength for the optical transitions at 69 and 73 cm-1 was found, as it is shown in Fig. 5(a) for several values of magnetic field. The frequency of the corresponding excitations did not depend on the ramping history for magnetic field.

Figure 6(a) shows the transmission intensity map measured at 40 K for $H \parallel [111]$ with increments of 0.05 T. One can see that optical phonon at 81 cm$^{-1}$ remains unchanged, thus ruling out the possible magnetic-field-dependent contribution to the magnetodielectric effect. The oscillator strengths [Fig. 6(b)] for both hybrid modes decrease at the magnetic field of ~0.3 T. Similar drops for the oscillator strength dependencies have been observed for all measured temperatures between 5 and 70 K and for two directions of the applied magnetic field: [111] and [100]. This result is not expected since the static dielectric constant *increases* by a few percents in the same range of magnetic field. Figure 6(c) shows results from Ref. [4] for the change of the static dielectric constant in a weak magnetic field. Although there is a strong nonlinear shift for the LF excitation and a significant modification of the mode oscillator strength, the harmonic oscillator model cannot reproduce the experimental data for the magnetoelectric effect [Fig. 6(c)]. We can only make a speculative assumption that the combined oscillator strength of two hybrid LF-M modes redistributes, so that the increase of the magnetic part compensates for the decrease of the electric part. The observed variation of the oscillator strength for the hybrid modes can be related to the magnetostriction effect that takes place in exactly the same range of magnetic fields.

VI. CONCLUSIONS

We observed a strong hybridization between the magnon and LF transitions. This effect occurs when the corresponding excitations have comparable frequencies. The magnon in Tb-IG is confined between two LF transitions: it couples to the higher-energy one at low temperatures and weak fields, while it comes to a resonance with the low-energy LF transition at higher temperatures of about 60 K. The corresponding coupling energies are 2 and 6 cm$^{-1}$. Redistribution of the oscillator strength between these IR excitations is a consequence of their strong hybridization. The temperature-induced variation of the static dielectric constant seems to correlate well with the softening and the oscillator strength variation of the hybrid modes. The possible role of the rhombohedral distortions in this compound is in a reduction of symmetry on the $Tb^{3+}$ ion that results in the forced electric dipole activity of the LF optical transitions. More detailed theoretical studies are required to confirm a possible connection between the hybrid modes and the magnetodielectric effect in Tb-IG.

The authors are thankful to S. M. O'Malley, L. Mihály, and T. Zhou for valuables discussions and to R. Smith for help at U4IR and U12IR beamlines. T. D. K. and E. S. at NJIT were supported by the NSF-DMR-0546985. V. K. and S-W. C. at Rutgers were supported by DE-FG02-07ER46382. Use of the National Synchrotron Light Source, Brookhaven National Laboratory, was supported by the U.S. Department of Energy, Office of Science, Office of Basic Energy Sciences, under Contract No. DE-AC02-98CH10886.

FIGURE CAPTIONS

FIG. 1 (color online) Normalized far-IR transmission spectra for a $Tb_3Fe_5O_{12}$ single crystal measured in a zero external magnetic field at $T = 5$, 17, and 80 K: (a), (b), and (c), respectively. The light propagation is along the [1 1 1] direction. Arrows indicate the frequencies of IR-active absorption lines. The weak intensity oscillations between 20 and 35 cm$^{-1}$ are the interference thickness fringes.

FIG. 2 (color online) (a) Maps of the normalized transmitted intensity *vs.* temperature and frequency for $Tb_3Fe_5O_{12}$. The blue (dark) color corresponds to stronger absorption, red (light) color indicates high transmission. The intensity scale is [0, 0.15]. (b) Experimental values for the ligand-field (LF) and the hybrid LF-M excitations. The results of the fit using Eq. (3) with the coupling constant $\Delta_{LF-M} = 6$ cm$^{-1}$ are shown with blue dashed curves. (c) |LF> and |M> wavefunction amplitudes for the upper energy $\left| \Omega_{LF-M}^{(2)} \right\rangle$ hybrid state.

FIG. 3 (color online) (a) Temperature dependence for the oscillator strength for the two hybrid modes with the frequencies $\Omega_{LF-M}^{(1)}$ and $\Omega_{LF-M}^{(2)}$. In the temperature range between 60 and 80 K the modes are strongly coupled and only their sum can be determined. (b) Total contribution of the hybrid modes to the static values of $\varepsilon(0,T)$ and $\mu(0,T)$ calculated from the transmission spectra using Eq. (4). (c) The temperature-induced variation of the static dielectric constant for $Tb_3Fe_5O_{12}$ at zero magnetic field and at the magnetic field of 0.2 T (from Ref. [4]).

FIG. 4 (color online) (a) Maps of the normalized transmitted intensity *vs.* magnetic field and frequency for $Tb_3Fe_5O_{12}$ at $T$=15 K and $H \parallel [\ 1\ 1\ 1\ ]$. The blue (dark) color corresponds to stronger absorption, while red (light) color indicates high transmission. The scale of the transmission intensity is the same as that for the graphs in Figure 2(a). (b) Variation of the LF, magnon, and phonon excitations in magnetic field $H \parallel [\ 1\ 1\ 1\ ]$. The linear slopes for magnetic field dependence for $\Omega_{LF-M}^{(1)}$, $\Omega_{LF-M}^{(2)}$, and $\Omega_{LF-M}^{(3)}$ excitations are $1\mu_B$, $-1.5\mu_B$, and $7\mu_B$, respectively.

FIG. 5 (color online) (a) Normalized far-IR transmission spectra for a $Tb_3Fe_5O_{12}$ single crystal measured at $T$= 5 K for external magnetic field $H$ = 0, 0.3, 0.6, and 0.9 T. Both, the light propagation and magnetic field directions are along [1 0 0]. Arrows indicate the frequencies of two hybrid modes: $\Omega_{LF-M}^{(2)}$ and $\Omega_{LF-M}^{(3)}$. (b) Variation of the coupled LF-M excitations $\Omega_{LF-M}^{(2)}$ and $\Omega_{LF-M}^{(3)}$ in magnetic field $H \parallel [\ 1\ 0\ 0\ ]$. Blue solid lines are fits using Eq. (5) for coupled excitations. Dashed lines are strong-field approximations for uncoupled excitations.

FIG. 6 (color online) (a) Maps of the normalized transmitted intensity *vs.* magnetic field $H \parallel [\ 1\ 1\ 1]$ measured at $T$ = 40 K. The scale of intensity is [0, 0.1]. (b) Results of the fit for the oscillator strengths for two low-frequency hybrid modes: $\Omega_{LF-M}^{(1)}$ and $\Omega_{LF-M}^{(2)}$. Dashed curves guide the eye. (c) Magnetic field dependence of the dielectric constant at various $T$ from Ref. [4].

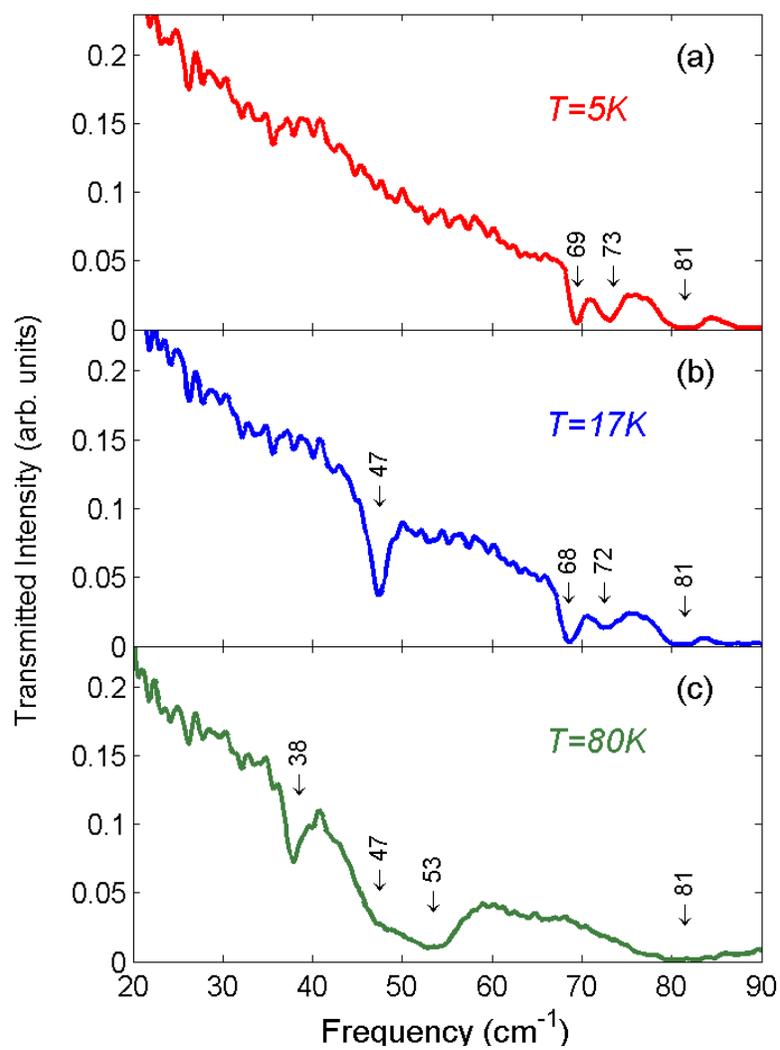

FIG. 1

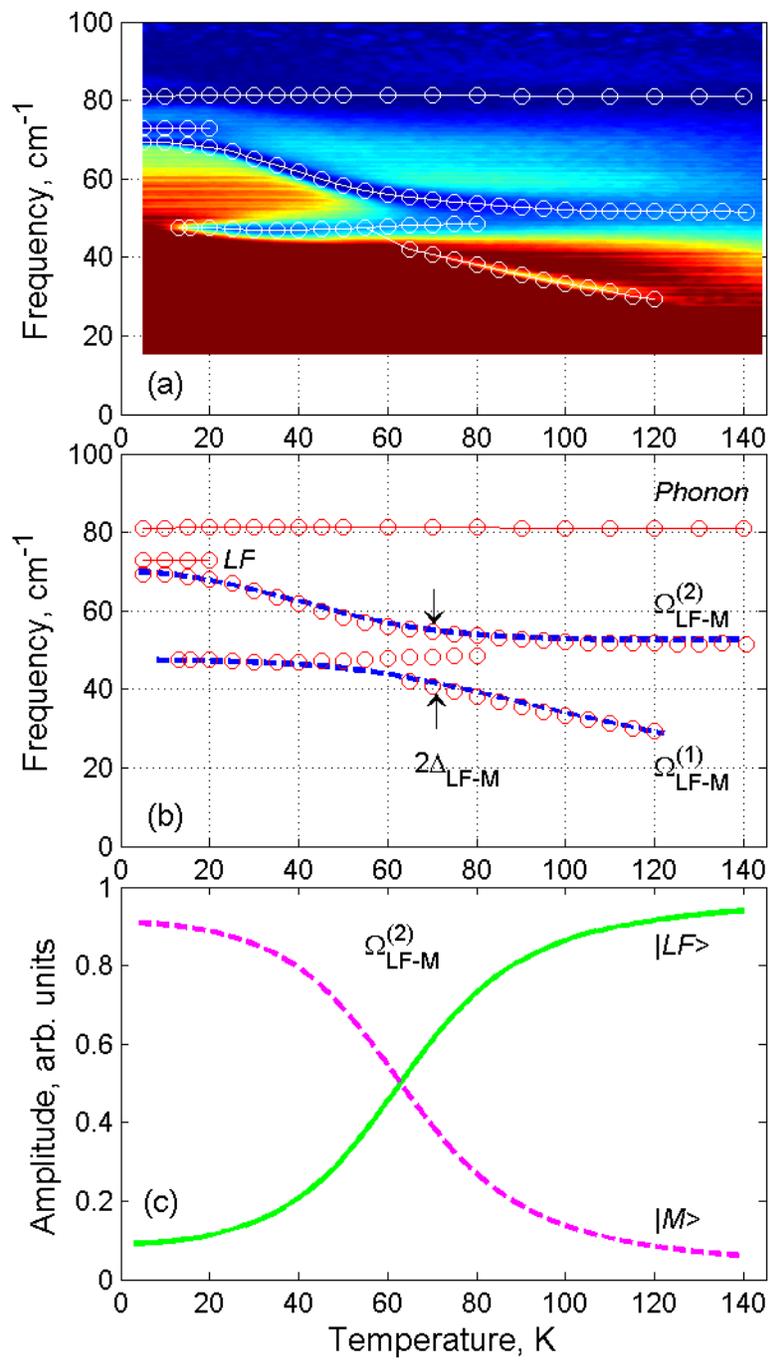

FIG.2

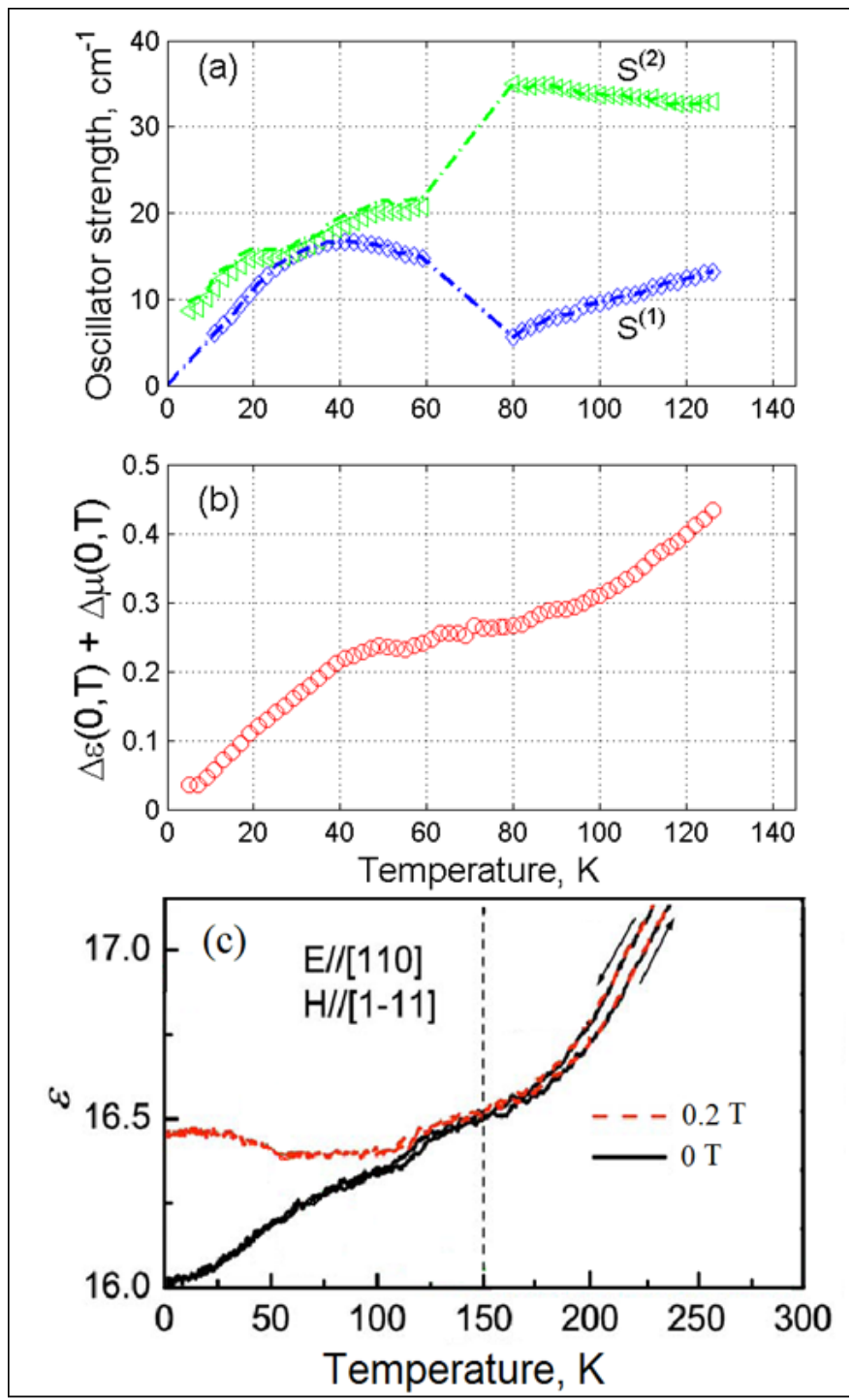

FIG. 3

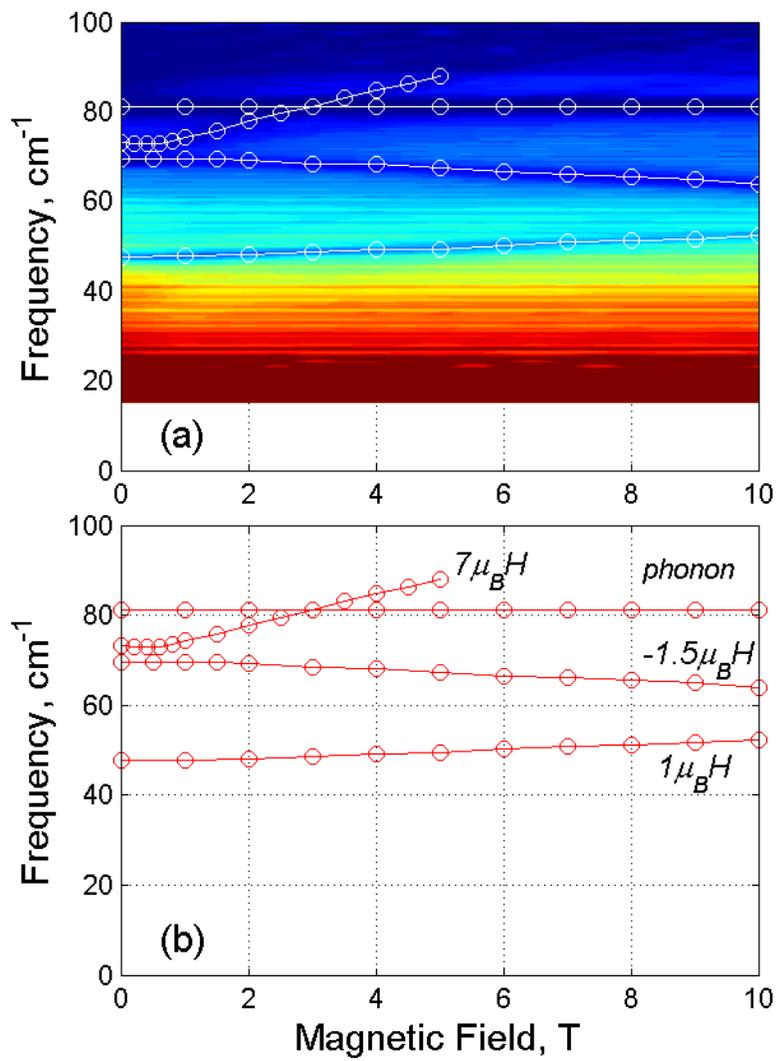

FIG. 4

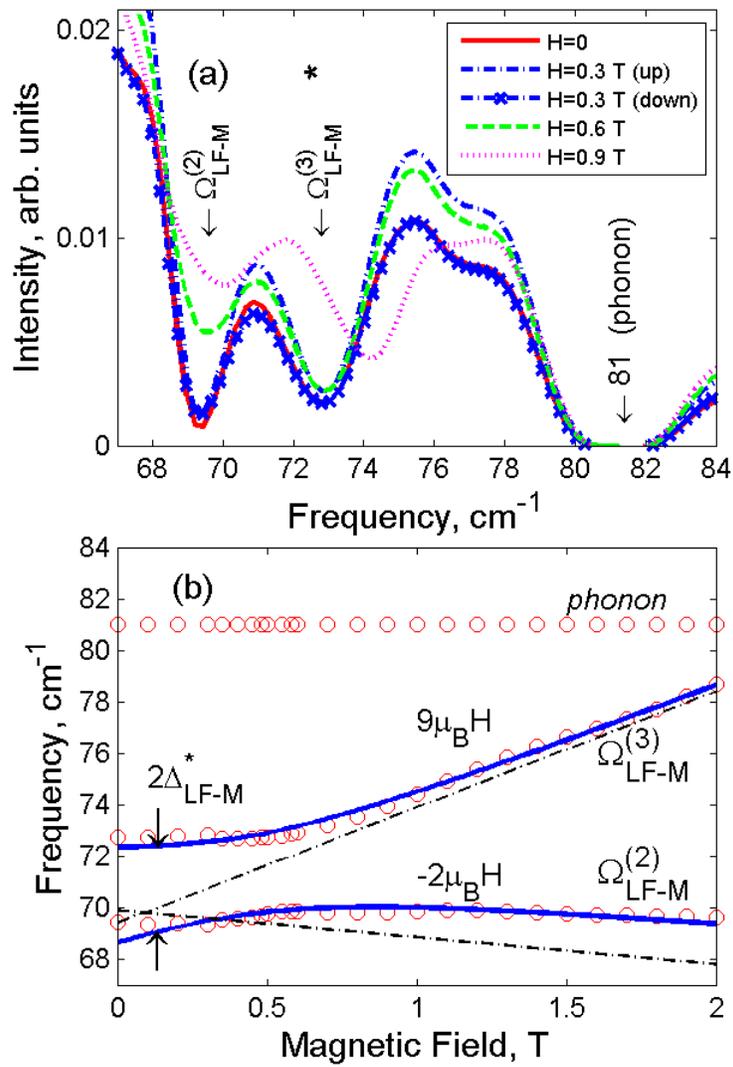

FIG. 5

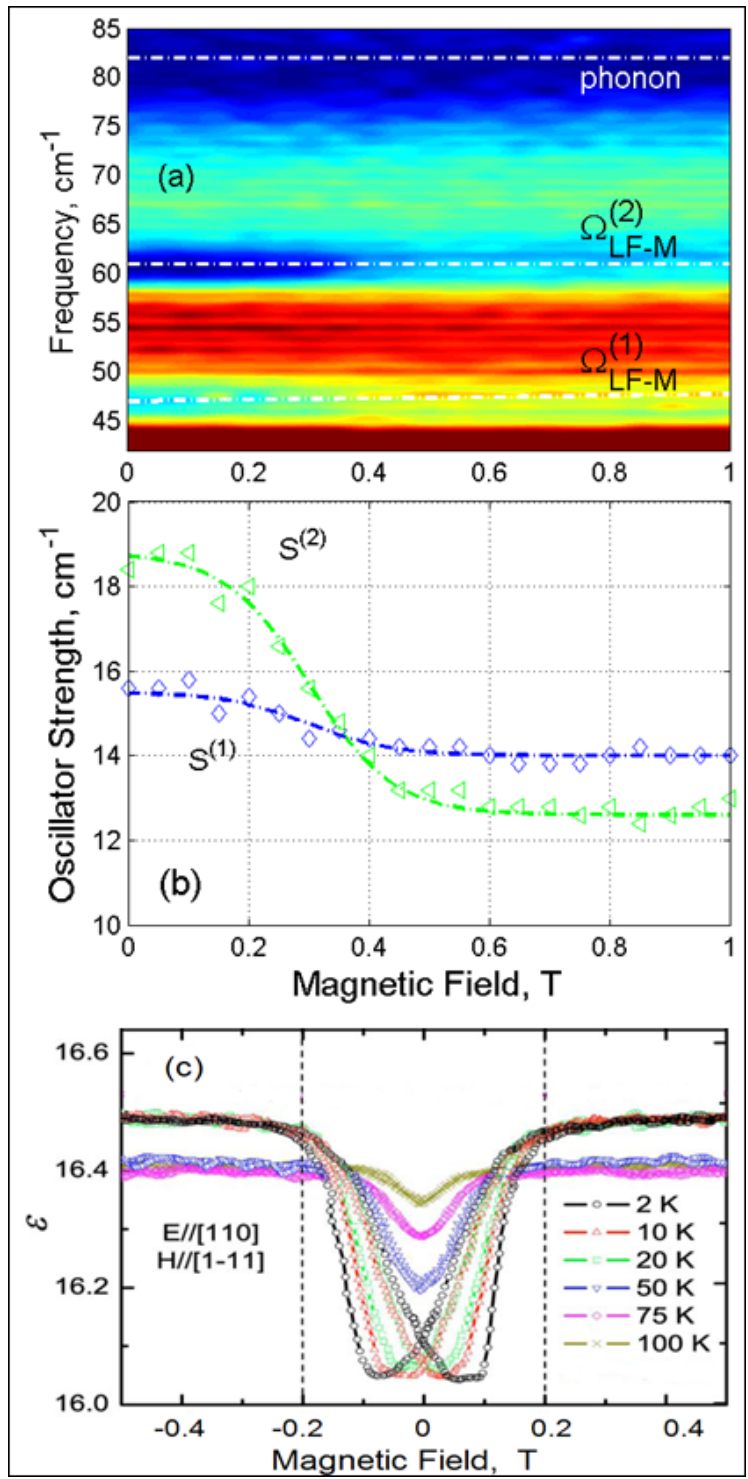

FIG. 6